\documentclass[twocolumn,english,aps,prb,superscriptaddress]{revtex4-1}
\usepackage[T1]{fontenc}
\usepackage[latin9]{inputenc}
\usepackage{geometry}
\usepackage{refstyle}
\usepackage{amsmath}
\usepackage{amssymb}
\usepackage{graphicx}
\usepackage{wasysym}

\makeatletter

\AtBeginDocument{\providecommand\subsecref[1]{\ref{subsec:#1}}}
\providecommand{\tabularnewline}{\\}
\RS@ifundefined{subsecref}
  {\newref{subsec}{name = \RSsectxt}}
  {}
\RS@ifundefined{thmref}
  {\def\RSthmtxt{theorem~}\newref{thm}{name = \RSthmtxt}}
  {}
\RS@ifundefined{lemref}
  {\def\RSlemtxt{lemma~}\newref{lem}{name = \RSlemtxt}}
  {}

\usepackage{babel}

\usepackage{babel}

\makeatother

\usepackage{babel}
\begin{document}

\title{Inclusion--Exclusion Principle for Many-Body Diagrammatics}

\author{Aviel Boag}

\affiliation{School of Chemistry, Tel Aviv University, Tel Aviv 6997801, Israel}

\author{Emanuel Gull}

\affiliation{Department of Physics, University of Michigan, Ann Arbor, Michigan
48109, USA}

\author{Guy Cohen}

\affiliation{The Raymond and Beverley Sackler Center for Computational Molecular
and Materials Science, Tel Aviv University, Tel Aviv 6997801, Israel}

\affiliation{School of Chemistry, Tel Aviv University, Tel Aviv 6997801, Israel}
\begin{abstract}
Recent successes in Monte Carlo methods for simulating fermionic quantum impurity models have been based on diagrammatic resummation techniques, but are restricted by the need to sum over factorially large classes of diagrams individually. We present a fast algorithm for summing over the diagrams appearing in Inchworm hybridization expansions. The method relies on the inclusion--exclusion principle to reduce the scaling from factorial to exponential. We analyze the growth rate and compare with related algorithms for expansions in the many-body interaction. An implementation demonstrates that for a simulation of a concrete physical model at reasonable parameters and accuracy, our algorithm not only scales better asymptotically, but also provides performance gains of approximately two orders of magnitude in practice over the previous state-of-the-art. 
\end{abstract}
\maketitle

\section{Introduction}

The accurate description of systems of many strongly interacting fermions
is one of the big open problems in modern theoretical physics.\cite{leblanc_solutions_2015}
Apart from a few very special situations, all known exact and general
solutions scale exponentially in the number of degrees of freedom.
In order to make progress, approximate numerical methods that are both
precise and efficient enough to describe the salient aspects of the
problem need to be designed.

The solution of quantum impurity models, which describe a small interacting
region (an ``impurity'' or ``dot'') coupled to an infinite noninteracting
region (``leads'' or ``baths''), is much simpler than the general
problem but remains a formidable challenge.\cite{bulla_numerical_2008}
Quantum impurity models appear in a wide range of contexts, including
in the description of magnetic atoms embedded in a host material\cite{anderson_localized_1961}
or adsorbed on a surface,\cite{brako_slowly_1981} in the description
of quantum transport through mesoscopic systems\cite{kouwenhoven_electron_1997,datta_electronic_1997,goldhaber-gordon_kondo_1998,potok_observation_2007}
and molecules,\cite{aviram_molecular_1974,aviram_molecular_1998,nitzan_electron_2001,nitzan_electron_2003}
and in quantum embedding algorithms.\cite{georges_dynamical_1996,zgid_finite_2017}
Even greater challenges are faced where access to real-time dynamics
or the description of high-lying excitations is needed. Numerical
methods that are able to describe these phenomena reliably and efficiently
are therefore highly desired.

The stochastic sampling of terms in a many-body perturbation theory,
known as ``diagrammatic''\cite{prokofev_polaron_1998} or ``continuous-time''\cite{gull_continuous-time_2011,rubtsov_continuous-time_2005,werner_continuous-time_2006,werner_efficient_2007,gull_continuous-time_2008}
quantum Monte Carlo, has been highly successful at describing the
equilibrium physics of impurity models. However, as systems are enlarged,
frustration is introduced, or equations are generalized to real-time
propagation,\cite{muhlbacher_real-time_2008,werner_diagrammatic_2009,werner_weak-coupling_2010,schiro_real-time_2009,schiro_real-time_2010}
the straightforward formulation of these algorithms scales exponentially
due to either the fermionic or the dynamical sign problem. This motivates
the need for formulations that either eliminate this exponential scaling
entirely or delay its onset for long enough that useful results can
be obtained with available resources.

Several such attempts have been made for lattice models.\cite{prokofev_bold_2007,prokofev_bold_2008,prokofev_diagrammatic_2013}
They are based on using the underlying structure of many-body diagrammatics
to reduce the number of diagrams that need to be considered, \textit{e.g.}
by considering connected diagrams only in a Green's function series,
by considering irreducible diagrams only in a self-energy expansion,
or by employing the ``skeleton'' technique to self-consistently
resum (or ``boldify'') certain classes of diagrams. These techniques
typically trade an alleviated sign problem (caused by the reduction
of the number of diagrams) against increased algorithmic complexity
and, potentially, convergence issues.\cite{kozik_nonexistence_2015}

In the context of impurity models, these techniques have mostly found
application in the Keldysh diagrammatics for real-time propagation.\cite{gull_numerically_2011,cohen_numerically_2013,cohen_greens_2014,cohen_greens_2014-1}
In a first implementation, partial summations (boldification) based
on semi-analytic impurity model techniques\cite{gull_bold-line_2010,cohen_greens_2014,cohen_greens_2014-1}
could substantially alleviate the sign problem, and in some cases
allow evaluation of slow dynamics.\cite{cohen_memory_2011,cohen_numerically_2013,cohen_generalized_2013}
Later, the realization that the causal structure of real-time dynamics
could be integrated directly into the algorithm led to the so-called
Inchworm method,\cite{cohen_taming_2015} which for several systems
and expansions seems to overcome the dynamical sign problem entirely
or in a wide range of physical regimes.\cite{antipov_currents_2017,dong_quantum_2017,chen_inchworm_2017,chen_inchworm_2017-1,ridley_numerically_2018}

However, all of these methods rely on an explicit enumeration of all
allowed diagrams at a given set of $n$ perturbation times for diagrams
of order $n$. This enumeration is expensive, since it scales as $n!$.
Access to large diagram order is therefore prohibitively expensive,
and the applicability of the various methods is restricted to domains
where convergence is obtained at relatively small orders.

In this paper we present a method that replaces the explicit enumeration
of $n!$ diagrams in the Inchworm hybridization expansion with a fast
summation algorithm based on the inclusion--exclusion principle.
We develop theoretical bounds for the scaling of the algorithm and
describe results from a practical implementation. We also compare
our method to a reformulation of the diagram summation in the interaction
expansion,\cite{rossi_polynomial_2017,rossi_determinant_2017} showing
that while our method is superior in the context of hybridization
expansions, the method of Ref.~\onlinecite{rossi_determinant_2017}
remains superior in the context of interaction expansions.

The remainder of this paper proceeds as follows: in Sec.~\ref{sec:Method},
we define the necessary concepts and then present our inclusion--exclusion
algorithm for the hybridization expansion, as well as two optimizations.
The algorithm of Ref.~\onlinecite{rossi_determinant_2017} for the interaction expansion
is reviewed, and a hybridization-expansion algorithm along similar lines is presented and compared to the inclusion--exclusion algorithm. An inclusion--exclusion
algorithm for the interaction expansion is then presented and compared
to that of Ref.~\onlinecite{rossi_determinant_2017}. Sec.~\ref{sec:Results} includes first a direct
comparison of the direct and inclusion--exclusion summation methods,
then a comparison of their performance within the Inchworm algorithm
for population dynamics in an Anderson impurity model. In Sec.~\ref{sec:Conclusions}
we conclude. Two appendices are also provided: appendix~\ref{sec:Derivation}
presents a derivation of our main formula from the inclusion--exclusion
principle, and appendix~\ref{sec:Computational-efficiency} presents
the methodology behind the theoretical expressions for the computational
scaling of the algorithms and optimizations we discuss.

\section{Method\label{sec:Method}}

The standard continuous-time hybridization expansion (``bare'' CTHYB)
in imaginary time\cite{werner_continuous-time_2006} and real time\cite{muhlbacher_real-time_2008,werner_diagrammatic_2009,schiro_real-time_2009}
has been described in the literature, and we refer readers interested
in the details of the expansion to previous work. For the purposes
of the present work, it is sufficient to introduce a simplified description
of the diagrammatic structure and the process of evaluating diagrams.
As the main idea we wish to present is general, we will do this in
a form that is largely agnostic to the details of the model. Furthermore,
in order to provide a self-contained description of the algorithm
introduced in this paper, we will also introduce a few concepts from
the Inchworm CTHYB expansion; once again, for a full discussion readers
are referred to the existing literature.\cite{cohen_taming_2015,antipov_currents_2017}

\subsection{Definitions}

Consider a generic impurity model Hamiltonian split into two parts:
\begin{equation}
\hat{H}=\hat{H}_{0}+\hat{V}.
\end{equation}
Here, $\hat{H_{0}}=\hat{H}_{D}+\hat{H_{B}}$ is separated into ``dot''
and ``bath'' subspaces, the second of which is noninteracting (\emph{i.e.}
described by a quadratic Hamiltonian); and $\hat{V}$ is a hybridization
Hamiltonian connecting the two subspaces. We assume that every element
in the Hamiltonian can be written in terms of second quantization
operators $\hat{a}_{k}$ and $\hat{a}_{k}^{\dagger}$ obeying fermionic
commutation relations, with $k$ enumerating the degrees of freedom.
The time dependence of the expectation value of some observable $\hat{A}$
is then given by 
\begin{equation}
\left\langle \hat{A}\left(t\right)\right\rangle =\left\langle \hat{U}^{\dagger}\left(t\right)\hat{A}_{I}\left(t\right)\hat{U}\left(t\right)\right\rangle ,\label{eq:time_dependent_observable}
\end{equation}
where for any operator $\hat{O}$, $\hat{O}_{I}\left(t\right)\equiv e^{i\hat{H}_{0}t}\hat{O}e^{-i\hat{H}_{0}t}$,
and $\left\langle \ldots\right\rangle $ signifies a trace on all
degrees of freedom with respect to some initial density matrix. The
interaction picture propagator $\hat{U}\left(t\right)\equiv e^{i\hat{H}_{0}t}e^{-i\hat{H}t}$
can be written in the form 
\begin{equation}
\begin{aligned}\hat{U}\left(t\right) & =\sum_{n=0}^{\infty}\left(-i\right)^{n}\int_0^t\mathrm{d}t_{1}\cdots\int_0^{t_{n-1}}\mathrm{d}t_{n}\\
 & \times\hat{V}_{I}\left(t_{1}\right)\cdots\hat{V}_{I}\left(t_{n}\right).
\end{aligned}
\label{eq:interaction_picture_propagator}
\end{equation}
In diagrammatic Monte Carlo techniques, the high-dimensional time
integrals appearing when Eq.~(\ref{eq:interaction_picture_propagator})
is replaced into Eq.~(\ref{eq:time_dependent_observable}) are carried
out stochastically by sampling the times at which the $V_{I}\left(t\right)$,
called \emph{vertices}, appear. This requires that we be able to efficiently
evaluate the integrands
\begin{equation}
\left\langle \hat{V}_{I}\left(t_{1}\right)\cdots\hat{V}_{I}\left(t_{n}\right)\hat{A}_{I}\left(t\right)\hat{V}_{I}\left(t_{1}^{\prime}\right)\cdots\hat{V}_{I}\left(t_{n}^{\prime}\right)\right\rangle ,
\end{equation}
where the times $0<t_{i},t_{i}^{\prime}<t$ come from terms in Eq.~(\ref{eq:time_dependent_observable})
for $\hat{U}^{\dagger}\left(t\right)$ and $\hat{U}\left(t\right)$.

\begin{figure}
\includegraphics{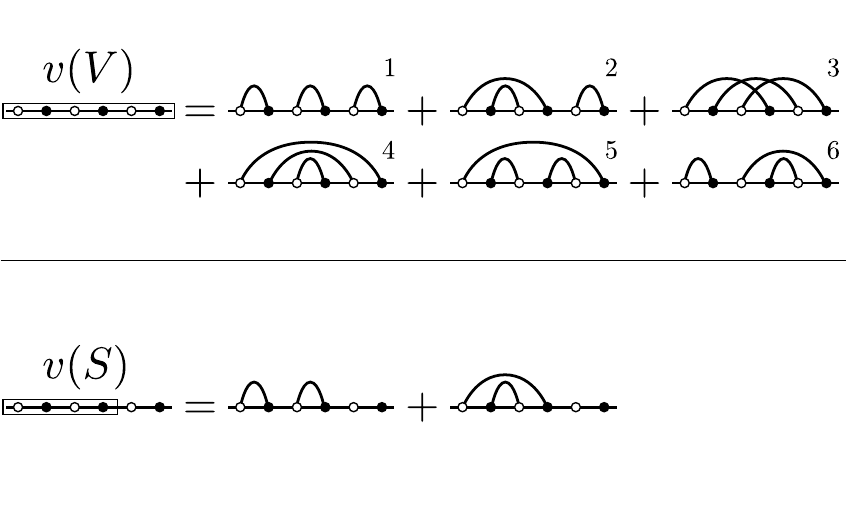}\caption{Elements of the bare hybridization expansion. The top panel shows
the diagrammatic representation of the sum $v\left(V\right)$, marked
by a box, over all diagrams generated by the complete set of vertices
at all times $V$. The vertices are denoted by filled and empty circles,
which indicate whether they consist of creation or annihilation operators,
respectively. In each diagram, every creation operator is connected
by a curved hybridization line to an annihilation operator. The bottom
panel shows the partial sum $v\left(S\right)$ over all diagrams generated
by a subset of four vertices, $S\subset V$.\label{fig:elements_bare}}
\end{figure}

In bare CTHYB, Eq.~(\ref{eq:time_dependent_observable}) is finally
written as 
\begin{equation}
\begin{aligned}\left\langle \hat{A}\left(t\right)\right\rangle  & =\sum_{n=0}^{\infty}\sum_{\left\{ s_{1},\ldots,s_{2n}\right\} }v\left(\left\{ s_{1},\ldots,s_{2n}\right\} \right)\\
 & \times p\left(\left\{ s_{1},\ldots,s_{2n}\right\} \right).
\end{aligned}
\end{equation}
Here, $p$ is a local propagator part that can be obtained from exact
diagonalization of the isolated dot Hamiltonian; and $v$, which is called
the lead influence functional, takes the form 
\begin{equation}
\begin{aligned}v\left(\left\{ s_{1},\ldots,s_{2n}\right\} \right) & \equiv\sum_{k_{1},\ldots,k_{2n}\in B}\gamma_{k_{1}}\gamma_{k_{2}}^{*}\cdots\gamma_{k_{2n-1}}\gamma_{k_{2n}}^{*}\times\\
 & \left\langle \hat{a}_{I,k_{1}}^{\dagger}\left(s_{1}\right)\hat{a}_{I,k_{2}}\left(s_{2}\right)\right.\\
 & \left.\cdots\hat{a}_{I,k_{2n-1}}^{\dagger}\left(s_{2n-1}\right)\hat{a}_{I,k_{2n}}\left(s_{2n}\right)\right\rangle_B .
\end{aligned}
\label{eq:hyb_influence_functional_definition}
\end{equation}
The $k$ indices are taken from the bath subspace only, and the $\gamma_{k}$
are parameters depending on the model. Averaging is performed only over the isolated bath subspace. Additional model-dependent
local indices which may appear in the expansion have been suppressed
for brevity.

The $s_{i}\in\left\{ t_{j},t_{j}^{\prime},t\right\} $ in Eq.~(\ref{eq:hyb_influence_functional_definition})
are a set of contour times, and together they are called a \emph{configuration}.
Since the part of $\hat{H}_{0}$ which includes bath operators is
quadratic, Eq.~(\ref{eq:hyb_influence_functional_definition}) can
be evaluated using Wick's theorem.\cite{negele_quantum_1998} This
results in a sum over $n!$ different diagrams, each of which corresponds
to a permutation of $n$ indices matching each creation operator with
an annihilation operator. Nevertheless, for fermions, this sum can
be evaluated at a computational cost which is cubic in $n$ because
it takes the form of a determinant:\cite{werner_continuous-time_2006,muhlbacher_real-time_2008}
\begin{equation}
v\left(\left\{ s_{1},\ldots,s_{2n}\right\} \right)=\mathrm{Det}M\left(s_{1},\ldots,s_{2n}\right).\label{eq:hyb_influence_functional_determinant}
\end{equation}
The elements of the matrix $M$ are given by a set of interaction
picture correlation functions which can be easily evaluated, since
they describe time evolution within a noninteracting reference system:
\begin{equation}
\begin{aligned}M_{ij} & =\sum_{k_{2i+1},k_{2j}}\gamma_{k_{2i+1}}\gamma_{k_{2j}}^{*}\\
 & \times\left\langle \hat{a}_{I,k_{2i+1}}^{\dagger}\left(s_{2i+1}\right)\hat{a}_{I,k_{2j}}\left(s_{2j}\right)\right\rangle_B .
\end{aligned}
\label{eq:hybridization_matrix}
\end{equation}

The top panel of Fig.~\ref{fig:elements_bare} illustrates the connection
between determinants and diagrams.\cite{werner_continuous-time_2006}
The determinant of Eq.~(\ref{eq:hyb_influence_functional_determinant})
is represented by a box, with filled (empty) circles representing
the times at which creation (annihilation) operators appear in a particular
configuration. We have chosen a certain $6^{\mathrm{th}}$ order (\emph{i.e.}
the perturbation order $2n=6$ or $n=3$) configuration. The terms
comprising the determinant, each of which corresponds to a particular
permutation pairing the $n$ creation operators to the $n$ annihilation
operators, delineate $n!=6$ individual diagrams. The so-called hybridization
lines in the diagrams signify pairings, and a line between operators
at times $s_{2i}$ and $s_{2j+1}$ corresponds to a multiplicative
factor of $M_{ij}$ from Eq.~\ref{eq:hybridization_matrix}. We denote
the sum over all diagrams generated by the complete set of vertices
$V$, with $\left|V\right|=2n$, as $v\left(V\right)$. In the lower
panel of of Fig.~\ref{fig:elements_bare}, we show a sum over all
diagrams generated by some $S\subset V$, which can also be evaluated
as a determinant.

The value of each diagram is a product of the hybridization functions
of Eq.~(\ref{eq:hybridization_matrix}) multiplied by an additional
fermion sign determined by the choice of permutation, or equivalently
a term in Eq.~(\ref{eq:hyb_influence_functional_determinant}); and
by the local propagator $p\left(\left\{ s_{1},\ldots s_{2n}\right\} \right)$
which does not depend on the permutation and is therefore not of interest
in the present context. The fermion sign is suppressed in our diagrammatic
notation for clarity, but is crucial in order for the sum to form
a determinant.

The bare CTHYB expansion of Ref.~\onlinecite{werner_continuous-time_2006}
benefits greatly from this determinant structure and the resulting
polynomial cost of evaluating the sum of all diagrams associated with
a configuration. Essentially, it means that time configurations rather
than individual diagrams form the sampling space. However, the real
time bare expansions,\cite{muhlbacher_real-time_2008,werner_diagrammatic_2009,werner_weak-coupling_2010,schiro_real-time_2009,schiro_real-time_2010}
as well as their bold counterparts,\cite{gull_bold-line_2010,gull_numerically_2011,cohen_greens_2014,cohen_greens_2014-1}
suffer from a dynamical sign problem: as the propagation time $t$
increases, the stochastic error increases exponentially.

\subsection{Fully connected, $k$-connected, proper and improper diagrams}

The Inchworm algorithm overcomes the dynamical sign problem (in at
least some cases) by taking advantage of the causal diagrammatic properties
of the expansion and the fact that evaluating propagation over short
time intervals is numerically inexpensive.\cite{cohen_taming_2015}
However, this comes at a cost: within the Inchworm expansion, contributions
are written in terms of dressed propagators, and the sum over diagrams
for a particular configuration can no longer be written in the determinant
form of Eq.~(\ref{eq:hyb_influence_functional_determinant}).

It is therefore necessary to explicitly iterate over a factorial number
of permutations for each configuration and filter a subset of dressed
diagrams, which is typically still factorial. One then sums over the
factorial number of contributions corresponding to this subset individually,
resulting in an overall \emph{$O\left(n!\right)$} computational scaling
in the expansion order $2n$, which should be compared to \emph{$O\left(n^{3}\right)$
}scaling in bare expansions. Nevertheless, while the order needed
to converge bare expansions always increases with time, Inchworm expansions
can often be terminated at low orders. In such cases, the loss of
the determinant structure may be worthwhile, as the exponential scaling
in time due to the dynamical sign problem is removed.

\begin{figure}
\includegraphics{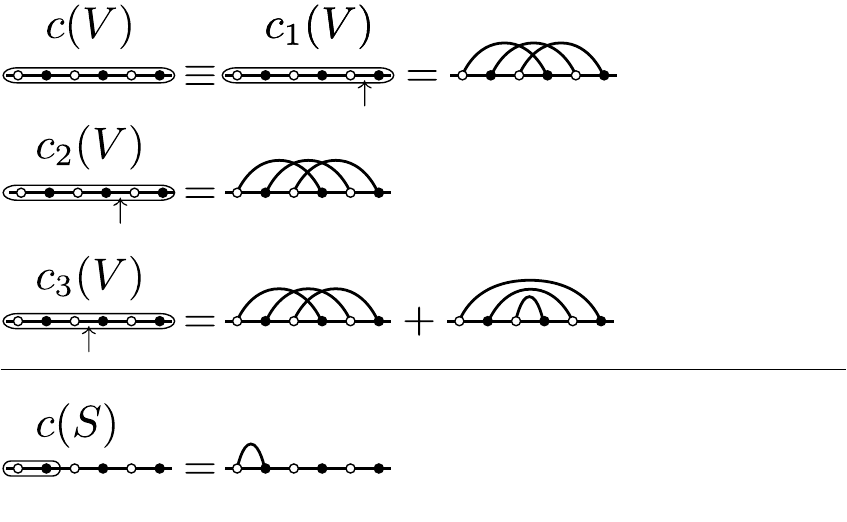}\caption{Elements of the Inchworm hybridization expansion. The top panel shows
the sum over all connected or $k$-connected diagrams generated by
the complete set of vertices $V$. This is denoted by a rounded box,
with an arrow delineating the value of $k$, \emph{i.e.} the boundary between proper and improper vertices. The connected and 1-connected
sums are identical by definition and include a single diagram; additionally,
for this particular case, no new 2-connected diagrams exist and one
more 3-connected diagram exists. The lower panel shows a sum over
connected diagrams generated by $S$, a subset of $V$ containing
two vertices.\label{fig:elements_inchworm}}
\end{figure}

In order to explain precisely which diagrams must be summed within
the Inchworm method, we first introduce the concept of connected and
$k$-connected diagrams. A diagram is considered (fully) connected
if all hybridization lines within it are connected by crossing. Note
that this differs from the more standard definition of connectedness
encountered in interaction expansions, where connectivity is a property
of the graph of vertices, which are connected by Green's function
lines. Here, it can be thought of as a property of the graph comprising
hybridization lines as nodes, with edges drawn between any two hybridization
lines which cross each other. Connectivity is illustrated in the top
panel of Fig.~\ref{fig:elements_inchworm}, where the sum $c\left(V\right)$
over connected diagrams generated by vertices $V$ is denoted by a
rounded box. Of the six diagrams in the top panel of Fig.~\ref{fig:elements_bare},
only diagram 3 is connected.

A diagram is $k$-connected if each of its connected components contains
at least one of $k$ special vertices, which we refer to as \emph{improper},
whereas the other $2n-k$ vertices are termed \emph{proper}. In the
diagrams discussed here, improper vertices are always the rightmost
$k$ vertices. $1$-connectedness is identical to full connectedness,
since there can be only one connected component containing the single
improper vertex. In the upper panel of Fig.~\ref{fig:elements_inchworm},
the sum $c_{k}\left(V\right)$ over $k$-connected diagrams generated
by all vertices $V$ is denoted by a rounded box with an arrow to
the left of the $k$ improper vertices. For the particular configuration
we have chosen to discuss, the sum $c_{1}\left(V\right)$ over 1-connected
and $c_{2}\left(V\right)$ over 2-connected diagrams is the same,
but the sum $c_{3}\left(V\right)$ over 3-connected diagrams contains
one additional term corresponding to diagram 4 in the top panel of Fig.~\ref{fig:elements_bare}.

In analogy to Fig.~\ref{fig:elements_bare}, the bottom panel shows
that it is also possible to define a sum over all connected diagrams
generated by a subset $S$ of the vertices $V$. In this case, we
chose a two-vertex subset which contains only a single diagram.

In the most basic Inchworm expansion,\cite{cohen_taming_2015} one
extends a known propagator over some time interval $\left(t_{i},t_{\uparrow}\right)$
into a longer propagator over the interval $\left(t_{i},t_{f}\right)$
with $t_{f}>t_{\uparrow}$. We set vertices in the interval $\left(t_{\uparrow},t_{f}\right)$
to be improper, and all other vertices to be proper. Given that there
are $k$ improper vertices, the mathematical problem that needs to
be addressed within the algorithm may then be reduced to the summation
of all $k$-connected diagrams.

\subsection{Application of the inclusion--exclusion principle\label{subsec:Naive-algorithm}}

\begin{figure}
\includegraphics{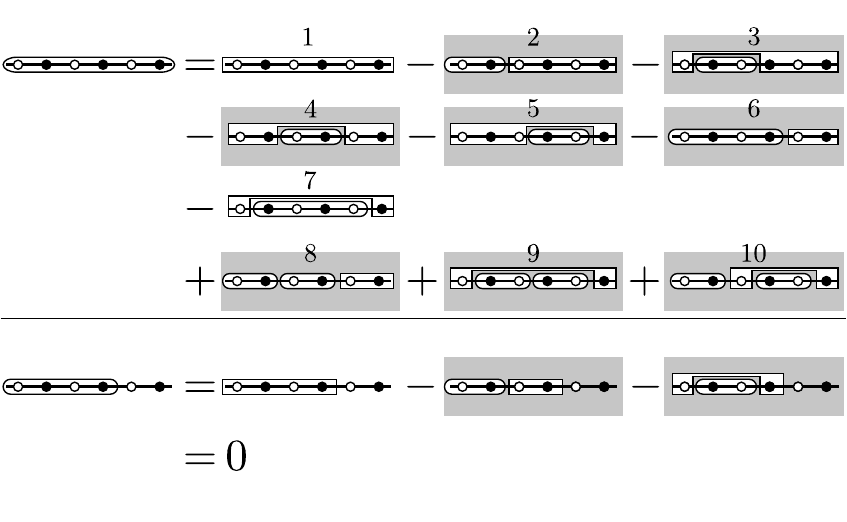}\caption{Illustration of the inclusion--exclusion algorithm. In the top panel,
the expansion for the sum over fully connected ($k=1$) diagrams for a $6^{\mathrm{th}}$ order
configuration is written in diagrammatic form. Below, the expression
for a particular element within this expansion is shown. Terms inside
shaded rectangles are removed by the optimizations.\label{fig:Basic-algorithm}}
\end{figure}

The inclusion--exclusion principle can be used to avoid the explicit
summation over a factorial number of $k$-connected diagrams. To see
how this works, we will first consider fully connected diagrams for
the same example configuration considered above (see the top panel
of Fig.~\ref{fig:Basic-algorithm}). Every disconnected diagram contains
at least one disconnected piece composed of lines fully spanning an adjacent
subset of the proper vertices. We will refer to an adjacent subset
as a \emph{segment}. Therefore, to obtain the set of connected diagrams,
one might start from the sum over all diagrams $v\left(V\right)$,
calculated as a determinant in polynomial time, and subtract all terms
with connected subsegments of $V$. To do this, one could try to sum
over all possible segments, taking connected diagrams within the segment
and all diagrams outside it. Only segments with the same number of
creation and annihilation operators need be considered.

However, a diagram containing two disconnected pieces would be subtracted twice in this manner: once for the term in which the segment includes one piece, and once for the term where it includes the other.
To cancel out this double-counting, one should now add all such diagrams, by introducing terms corresponding
to all possible \emph{pairs} of segments. This argument could be repeated
indefinitely, leading to a mathematical structure analogous to the
one that results from attempting to express the size of a union of
$N$ sets by summing the sets and their intersections. The formal
mathematical connection with this concept, known as the inclusion--exclusion
principle, is presented in appendix~\ref{sec:Derivation}. Our expression
for the sum over $k$-connected diagrams is as follows: 
\begin{equation}
c_{k}\left(V\right)=\sum_{j=0}^{n-k}\left(-1\right)^{j}\sum_{\left\{ S_{i}\right\} }v\left(V\backslash\bigcup_{i=1}^{j}S_{i}\right)\prod_{i=1}^{j}c_{1}\left(S_{i}\right).\label{eq:algorithm_naive}
\end{equation}
Here, $j$ is the number of segments, and the summation is over all
possible segments comprising the $2n-k$ proper vertices. We note
that the expression is given in terms of the $c_{1}\left(S_{i}\right)$,
which can be recursively evaluated from it. While we will show several substantial optimizations, Eq.~(\ref{eq:algorithm_naive}) describes
the central result of this publication. The complete process is illustrated
in Fig.~\ref{fig:Basic-algorithm} for our $6^{\mathrm{th}}$ order
configuration, with the bottom panel illustrating the evaluation of
one of the elements appearing in the sum (which is in this case zero,
a fact that we will take advantage of soon).

At first glance, it is not clear that this approach holds any advantage:
in fact, in Fig.~\ref{fig:Basic-algorithm} we sum over 10 elements
rather than the 6 in Fig.~\ref{eq:algorithm_naive}, even before
taking into account the fact that we must also perform more summations
to obtain the various elements appearing in the expansion. However,
consider the scaling: the sum in eq.~(\ref{eq:algorithm_naive})
is over all sets of segments. Naively, to count them, one notes that
there are $2^{2n-k}$ ways to decide which vertices will be included
in segments (ignoring for a moment the differences between creation
and annihilation operators, which decrease this number). In the worst
case, if all $2n-k$ are chosen, the number of ways to construct segments
from this set is the number of compositions of $2n-k$, of which there
are $2^{2n-k-1}$; so, at worst, the summation should scale as $2^{4n-2k-1}$.
We must also compute each of the $c_{1}\left(S_{i}\right)$, each
of which should be no more expensive, but there is only a quadratic
number $\left(2n-k\right)$$\left(2n-k-1\right)$ of these. Given
that each step entails the calculation of a single determinant at
$O\left(n^{3}\right)$, even a rough estimate of the asymptotic computational
complexity is $O\left(n^{5}4^{n}\right)$, high but substantially
less than factorial. In fact, as we show in appendix~\ref{sec:Computational-efficiency},
a more careful calculation shows that the correct scaling $C_{n}$
in this case can be bounded from above by 
\begin{equation}
L_{n}^{U}=O\left(n^{3}\alpha^{2n}\right),\label{eq:cost_upper}
\end{equation}
with $\alpha\approx1.8019$. To simplify the calculation, this estimate
assumes that all operators can be paired with all other operators,
which results in an overestimate of the complexity. An alternative
assumption is that half the vertices are creation (annihilation) operators,
but they are arranged in arbitrary order. In this case it is possible
to calculate a cost averaged over the orderings. We term this estimate
\begin{equation}
L_{n}^{L}=O\left(n^{3}\beta^{2n}\right),\label{eq:cost_lower}
\end{equation}
where $\beta\le\alpha$. This is neither a strict upper bound nor
a lower one. However, it may be expected to function as an effective
lower bound, since one could suppose the computational complexity
in most models to be strongly influenced by the worst case ordering.
We find that $\beta\approx1.5072$ (see appendix~\ref{sec:Computational-efficiency}).

It is possible to generalize the algorithm to expansions where any
two vertices might be paired, such that there is no distinction between
creation and annihilation operators. In this case, the determinant
is replaced by a Pfaffian. Pfaffians, like determinants, can be computed
in polynomial time, and everything else in the algorithm remains essentially
unmodified. Furthermore, the worst-case scaling criterion of Eq.~(\ref{eq:cost_upper})
becomes exact. This generalization is of some interest from the mathematical
viewpoint, and might be considered the solution of a simpler, cleaner
problem. However, it is not immediately clear to us that it has utility
in the physical context. We will therefore not explore it further
here.

\subsection{Optimizations}

We can improve the algorithm further. In particular, for the example
used in the illustration, it is possible to drop all but two (slightly
modified) terms, such that the sum over connected diagrams can be
obtained from the evaluation of a single order 3 determinant and two
order 2 determinants which turn out to be zero. In general, however,
the computational cost will remain exponential in the number of vertices.
While the algorithm we have presented already produces an improvement
of the scaling to exponential, the actual exponent can be reduced
further with the aid of a few simple observations. This is of course
worthwhile, because it provides an additional exponential improvement
in performance.

\subsubsection{First optimization: adjacent segments}

First, note that we compute $v$ on the same subset $V\backslash\bigcup_{i}S_{i}$
for many different sets of segments $\left\{ S_{i}\right\} $, since
adjacent segments occupy the same vertices as their union. For example,
in Fig.~\ref{fig:Basic-algorithm} diagrams 6 and 8 share the same
determinant, as do diagrams 7 and 9. Therefore, we can regroup the
sum by first summing over non-adjacent segments, and then summing
over all possible divisions of a segment into adjacent subsegments
(which, once again, can be enumerated as compositions). In order to
do this, we first rewrite Eq.~(\ref{eq:algorithm_naive}) in the
following form:

\begin{equation}
\begin{aligned}c_{k}\left(V\right) & =\sum_{j=0}^{n-k}\sum_{\left\{ S_{i}\right\} }v\left(V\backslash\bigcup_{i=1}^{j}S_{i}\right)\prod_{i=1}^{j}\left[-c_{1}\left(S_{i}\right)\right]\\
 & =\sum_{\left\{ S_{i}\right\} }v\left(V\backslash\bigcup_{i}S_{i}\right)\prod_{i}\left(-c\left(S_{i}\right)\right).
\end{aligned}
\end{equation}
The summation now runs simultaneously over all sets of proper segments,
with no regard as to how many segments are in a set. Now, if we let
$a\left(S\right)$ be the sum of values of all partitions of a given
segment $S$,

\begin{equation}
a\left(S\right)\equiv\sum_{\left\{ D_{i}\right\} }\prod_{i}\left(-c\left(D_{i}\right)\right),\label{eq:a_s}
\end{equation}
where the $\left\{ D_{i}\right\} $ are all possible partitions of
$S$ into adjacent subsegments, we can write

\begin{equation}
c_{k}\left(V\right)=\sum_{\left\{ A_{i}\right\} }v\left(V\backslash\bigcup_{i}A_{i}\right)\prod_{i}a\left(A_{i}\right),\label{eq:first_optimization}
\end{equation}
where the $\left\{ A_{i}\right\} $ are all disjoint non-adjacent
segments comprising proper vertices.

$a\left(S\right)$ must now be evaluated for every possible segment
$S$. If we choose $S$ to be any segment $\left\{ v_{1},\dots,v_{j}\right\} $,
it is easy to see that for the case where the last segment is of length
$\ell$, the contribution to $a$ is $a\left(\left\{ v_{1},\dots,v_{j-\ell-1}\right\} \right)c\left(\left\{ v_{j-\ell},\dots,v_{j}\right\} \right)$.
Repeating this argument for all possible lengths $\ell$ then gives
all contributions:

\begin{equation}
\begin{aligned}a\left(\left\{ v_{1},\dots,v_{j}\right\} \right) & =-\sum_{\ell=1}^{j}a\left(\left\{ v_{1},\dots,v_{j-\ell-1}\right\} \right)\\
 & \times c\left(\left\{ v_{j-\ell},\dots,v_{j}\right\} \right).
\end{aligned}
\end{equation}

As we show in appendix~\ref{sec:Computational-efficiency}, the effect
of this reformulation is a reduction in the computational complexity
to $\alpha\approx1.618$ and $\beta\approx1.4142$.

\subsubsection{Second optimization: removing two-vertex segments\label{subsec:Second-optimization}}

For a second optimization, one need only note that a hybridization
line between two adjacent proper vertices never crosses any other
hybridization line, and therefore can't be a part of a $k$-connected
diagram. Given this, it is possible to eliminate the values of all
such lines by setting the corresponding elements of $M_{ij}$ to zero.
After doing so, there is no longer any need to consider segments of
length two, and the complexity improves to $\alpha\approx1.4432$
and $\beta\approx1.2676$.

Let us now revisit Fig.~\ref{fig:Basic-algorithm}. In the top panel,
with the second optimization, every term except 1 and 7 (\emph{i.e.}
all terms outlined by shaded rectangles) can be immediately dropped.
The first term is a third order determinant of a modified $M_{ij}$
in which some of the elements have been set to zero. The seventh term,
similarly to the sixth term in the bottom panel, is a second order
determinant of a similarly modified submatrix of $M_{ij}$, which
turns out to be zero. We therefore see that even for this $n=3$ example,
the optimized inclusion--exclusion method requires the computation
of fewer terms than the direct algorithm.

\subsection{Inverted algorithm}

Recently, an algorithm was found that allows for summing all connected
diagrams in interaction expansions in exponential rather than factorial
time.\cite{rossi_determinant_2017} It was shown that this leads to
polynomial complexity for evaluating thermodynamic quantities in certain
regimes,\cite{rossi_polynomial_2017} and was later extended to the
summation of irreducible diagrams.\cite{rossi_direct_2018,moutenet_determinant_2018,simkovic_determinant_2017}
Eq.~(\ref{eq:algorithm_naive}) is reminiscent of the main result
of Ref.~\onlinecite{rossi_determinant_2017}. Rephrased in a slightly
modified form for easy comparison with the expressions presented here,
Ref.~\onlinecite{rossi_determinant_2017} proposed the following
formula for the sum over all connected diagrams within an interaction
expansion for a Hubbard model: 
\begin{equation}
c\left(E,V\right)=v\left(E,V\right)-\sum_{S\subsetneq V}c\left(E,S\right)v\left(\emptyset,V\backslash S\right).\label{eq:rossi_result}
\end{equation}
Here, $V$ and $E$ are sets of internal and external vertices, respectively;
$v\left(A,B\right)$ is the sum over all (interaction) diagrams generated
by external vertices $A$ and internal vertices $B$ (given by a certain
determinant); and $c\left(A,B\right)$ is the sum over all connected
diagrams with external vertices $A$ and internal vertices $B$. This
result is seemingly much simpler than Eq.~(\ref{eq:algorithm_naive}):
there is no inclusion--exclusion hierarchy and the summation is terminated
at the level of single subsets rather than sets of subsets. However,
since there is a sum over subsets rather than segments, the resulting
computational scaling is $O\left(3^{n}\right)$, exponentially worse
than in our case. Inspired by this work, we set out to see if our
algorithm could be formulated in a similar way, and if any advantage
might be gained by this for either problem.

\subsubsection{Inverted algorithm for the hybridization expansion\label{subsec:Inverted-algorithm-for-hyb}}

Comparing Eqs.~(\ref{eq:algorithm_naive}) and (\ref{eq:rossi_result}),
if we let internal (external) vertices correspond to proper (improper)
vertices, the expression is inverted: while in Eq.~(\ref{eq:rossi_result})
the subtracted contributions are connected to the external part, in
Eq.~(\ref{eq:algorithm_naive}) they are disconnected from it. With
this in mind, it is possible to derive a different way of evaluating
$c_{k}\left(V\right)$, where the improper vertices are always enclosed
in a $k$-connected element: 
\begin{equation}
\begin{aligned}c_{k}\left(V\right) & =v\left(V\right)\\
 & -\sum_{\left\{ A_{i}\right\} \backslash\left\{ \emptyset\right\} }c_{k}\left(V\backslash\bigcup_{i}A_{i}\right)\prod_{i}v\left(A_{i}\right).
\end{aligned}
\label{eq:inverted_algorithm}
\end{equation}
Here, the summation is over all sets of one or more non-adjacent segments
comprising proper points. This is in much closer analogy to Eq.~(\ref{eq:rossi_result}).
It is even more similar to Eq.~(\ref{eq:first_optimization}), other
than in the signs and the reversal of roles between $c$ and $v$;
what appeared as the first optimization in the inclusion--exclusion
algorithm is necessary here for correctness.

The computational scaling of this algorithm is less than factorial,
but unfortunately remains higher than that of the inclusion--exclusion
algorithm: as discussed appendix~\ref{sec:Computational-efficiency},
it can be bound at $\alpha\approx2.1935$ and $\beta\approx1.7321$
as presented; the second optimization still applies to it, at which
point we obtain $\alpha\approx1.8718$, still higher than even the
unoptimized algorithm based on Eq.~(\ref{fig:Basic-algorithm}); and $\beta\approx1.4861$,
larger than the smallest $\alpha$ for the inclusion--exclusion case.
The inverted algorithm is therefore less suitable than the inclusion--exclusion
algorithm for the hybridization expansion.

We present a summary of the theoretical computational complexities
characterizing the different algorithms and optimizations in Table~\ref{complexity-table},
and refer the reader to appendix~\ref{sec:Computational-efficiency}
for details.

\begin{table}
\begin{tabular}{|c|c|c|}
\hline 
Optimization Level  & $\alpha$  & $\beta$\tabularnewline
\hline 
\hline 
Unoptimized, Eq.~(\ref{eq:algorithm_naive})  & 1.8019  & 1.5072\tabularnewline
\hline 
$1^{\mathrm{st}}$ optimization, Eq.~(\ref{eq:first_optimization})  & 1.6180  & 1.4142\tabularnewline
\hline 
$2^{\mathrm{nd}}$ optimization, \subsecref{Second-optimization}  & 1.4432  & 1.2676\tabularnewline
\hline 
Inverted algorithm, Eq.~\ref{eq:inverted_algorithm}  & 2.1935  & 1.7321\tabularnewline
\hline 
Inverted alg. with $2^{\mathrm{nd}}$ optimization  & 1.8718  & 1.4861\tabularnewline
\hline 
\end{tabular}\caption{\label{complexity-table}Theoretical complexity of the two proposed
algorithms for the hybridization expansion at different levels of
optimization. $O\left(n^{3}\alpha^{2n}\right)$ is an overestimating
simplification assuming all operators can be connected to each other,
while $O\left(n^{3}\beta^{2n}\right)$ provides an average cost assuming
the operators are randomly ordered and is most likely an underestimate
of the cost.}
\end{table}

\subsubsection{Inclusion--exclusion algorithm for the interaction expansion}

The algorithm of Sec.~\ref{subsec:Naive-algorithm} is substantially
more efficient than the one of Sec.~\ref{subsec:Inverted-algorithm-for-hyb},
which is reminiscent of the one in Ref.~\onlinecite{rossi_determinant_2017}.
It is intriguing to consider whether the inclusion--exclusion principle
might also be useful in the context of the interaction expansion.
On one hand this is a conceptually simpler problem, because there
is less mathematical structure to it; but on the other hand a computationally
harder one, because one must consider subsets of vertices rather than
segments.

It is easy to see that, as an alternative to Eq.~(\ref{eq:rossi_result}),
the sum over connected diagrams can be recast in the following form:
\begin{equation}
\begin{aligned}c\left(E,V\right) & =\sum_{j=0}^{\infty}\left(-1\right)^{j}\sum_{\left\{ S_{i}\right\} }v\left(E,V\backslash\bigcup_{i}S_{i}\right)\\
 & \times\prod_{i}c\left(\emptyset,S_{i}\right).
\end{aligned}
\label{eq:interaction_result}
\end{equation}
Here, we perform the summation over all possible sets of disjoint
subsets of internal vertices $\left\{ S_{i}\right\} \subset V$. This
is analogous to Eq.~(\ref{eq:algorithm_naive}).

Let us now consider the computational complexity of Eq.~(\ref{eq:interaction_result}).
The asymptotically dominant contribution in this case is not the evaluation
of the determinants, which is $O\left(n^{3}2^{n}\right)$ for $n=\left|V\right|+\left|E\right|$,
but the sum itself. The disjoint subsets of a set with $n$ elements
are known as its partitions. The sequence of numbers counting the
partitions of sets of increasing size are the Bell numbers $B_{n}$,
which are asymptotically bound by\cite{berend_improved_2010} 
\begin{equation}
B_{n}<\left(\frac{0.792n}{\ln\left(n+1\right)}\right)^{n}.
\end{equation}
This is better than factorial complexity, but worse than exponential
(as is the corresponding lower bond). Therefore, the inclusion--exclusion
algorithm is better than the brute force approach to the interaction
expansion, but not nearly as efficient as that of Ref.~\onlinecite{rossi_determinant_2017}.
Nevertheless, the inclusion--exclusion principle may turn out to
be of interest within interaction expansions with more detailed structure,
such as in cases where one neglects long-ranged correlations; and
may also turn out to be more amenable to fast update schemes. As this
is beyond the scope of the present work, we leave it for future study.

\section{Results\label{sec:Results}}

\subsection{Comparison with direct algorithm}

To analyze the algorithm, we will begin by considering the computational
cost of the summation itself, with no regard to any physical context.
This allows for a cleaner exploration of the scaling and for a well-defined
comparison with the theoretical exponents $\alpha$ and $\beta$.
For this purpose, we implemented a brute-force summation over 
over all $k$-connected permutations for a given configuration (``Direct algorithm'') and the inclusion--exclusion algorithm with both optimizations for performing the same task
(``Fast algorithm''). We applied these implementations
to all possible vertex configurations at different perturbation
orders. Importantly, we verified that the results given by the two algorithms
are identical within numerical accuracy in all cases. We also measured
the average evaluation time per configuration. While the absolute
value of this time depends on the implementation details and hardware,
the scaling with the perturbation order should be largely independent
of such details and can be explored systematically. The result depends
to some degree on the details of the model, which may feature symmetries
limiting the possible configurations; for the present purpose, we
assume no such symmetries, and we have found (not shown) that enforcing
symmetries has a relatively small quantitative effect on the results.

\begin{figure}
\includegraphics[width=8.6cm]{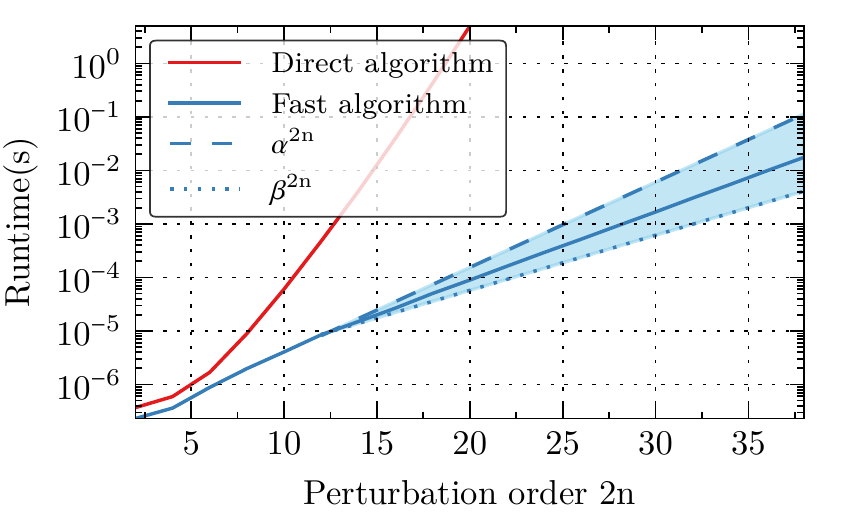}\caption{Comparison of runtimes for summing all 1-connected diagrams, using
implementations of both the direct algorithm and optimized inclusion--exclusion
algorithm. The results are averaged over all possible operator orderings
for a model without any symmetries. The theoretical upper bound $\alpha^{2n}$
and approximate lower bound $\beta^{2n}$ are also shown as dashed
and dotted lines, respectively, with the region between them shaded.\label{fig:algorithm_comparison_clean}}
\end{figure}

Fig.~\ref{fig:algorithm_comparison_clean} presents the average evaluation
time of the sum over all 1-connected diagrams as a function of the
perturbation order $2n$ (1-connected diagrams are the worst case
for our algorithm, and more general summations over the $k$-connected
diagrams appearing in the Inchworm expansion perform quantitatively,
if not qualitatively, better.). In comparison to the brute-force method, the inclusion--exclusion
algorithm exhibits superior scaling, which appears to be asymptotically
exponential as expected. The effective exponent is $\gamma\approx1.33$,
which lies below $\alpha$, as it must; and also lies above $\beta$.
We note that since we are averaging over operator orderings, the exponent
$\beta$ must be exact in the asymptotic limit, and any deviation
from it is due to the $n^{3}$ factor in the complexity Eq.~\ref{eq:cost_lower}.
This shows that the exclusion--exclusion algorithm works in practice:
it not only scales better than its direct counterpart, but also does
not feature a prohibitive prefactor that keeps it from being used
for small perturbation orders. In fact, the new algorithm appears
to always be faster, even at order 1.

We reiterate that the evaluation time per configuration shown in Fig.~\ref{fig:algorithm_comparison_clean}
is the average over possible operator orderings, as this more closely
reflects the use of the algorithm in a physical context. However,
it is also possible to consider the worst case. A similar analysis
(not shown) then leads to an exponent of $\gamma\approx1.4$, still
within the theoretical bounds but closer to the upper limit. We further
note that while asymptotically the factor $n^{3}$ in Eqs.~(\ref{eq:cost_upper})
and (\ref{eq:cost_lower}) becomes irrelevant, it is straightforward
to take it into account at finite $n$ using nonlinear function fitting.
While we verified that this procedure has a small quantitative effect
on the result, we did not use it in practice.

\subsection{Effect within Inchworm Monte Carlo}

Next, we consider what happens when we apply the inclusion--exclusion
algorithm to a concrete simulation of a physical model within Inchworm
Monte Carlo. We choose the Anderson impurity model addressed in the
original Inchworm paper:\cite{cohen_taming_2015} 
\begin{eqnarray}
H & = & \sum_{\sigma\in\left\{ \uparrow,\downarrow\right\} }\varepsilon d_{\sigma}^{\dagger}d_{\sigma}+Un_{\uparrow}n_{\downarrow}\label{eq:hamiltonian}\\
 &  & +\sum_{\sigma k}\varepsilon_{k}a_{\sigma k}^{\dagger}a_{\sigma k}+\sum_{a\sigma k}\left(\gamma_{k}a_{\sigma k}^{\dagger}d_{\sigma}+\mathrm{H.C.}\right).\nonumber 
\end{eqnarray}
Here, we set the dot's single-particle energy $\varepsilon$ to
$\varepsilon=-\frac{U}{2}$, where $U$ is the Hubbard interaction
energy, such that the system is particle--hole symmetric. The $d_{\sigma}$
and $d_{\sigma}^{\dagger}$ are dot fermionic annihilation and creation
operators, and the $a_{\sigma k}$ and $a_{\sigma k}^{\dagger}$ are
corresponding operators on the lead. The lead single-particle energies
$\varepsilon_{k}$ and the dot--lead hybridization terms $\gamma_{k}$
are determined so as to produce a flat band with overall coupling
strength $\Gamma$, cutoff energy $\Omega_{C}$ and cutoff width $\frac{1}{\nu}$:
\begin{eqnarray}
\begin{aligned}\Gamma\left(\omega\right) & \equiv2\pi\sum_{k}\gamma_{k}^{*}\gamma_{k}\delta\left(\omega-\varepsilon_{k}\right)\\
 & =\Gamma/\left[\left(1+e^{\nu\left(\omega-\Omega_{c}\right)}\right)\left(1+e^{-\nu\left(\omega+\Omega_{c}\right)}\right)\right].
\end{aligned}
\end{eqnarray}
Our choice of physical parameters will be motivated by our interest
in exploring a problem where high perturbation orders are important.
We will therefore arbitrarily select parameters which are particularly
difficult for the hybridization expansion. Throughout this work, we
set $U=3\Gamma$, $\Omega_{C}=100\Gamma$, $\nu\Gamma=10$. Additionally,
the inverse temperature of the bath is set to $\beta\Gamma=100$ and
its chemical potential is $\mu=0$. The dot is initially in the unoccupied
state and decoupled from the bath; at time zero the coupling is suddenly
activated.

\begin{figure}
\includegraphics[width=8.6cm]{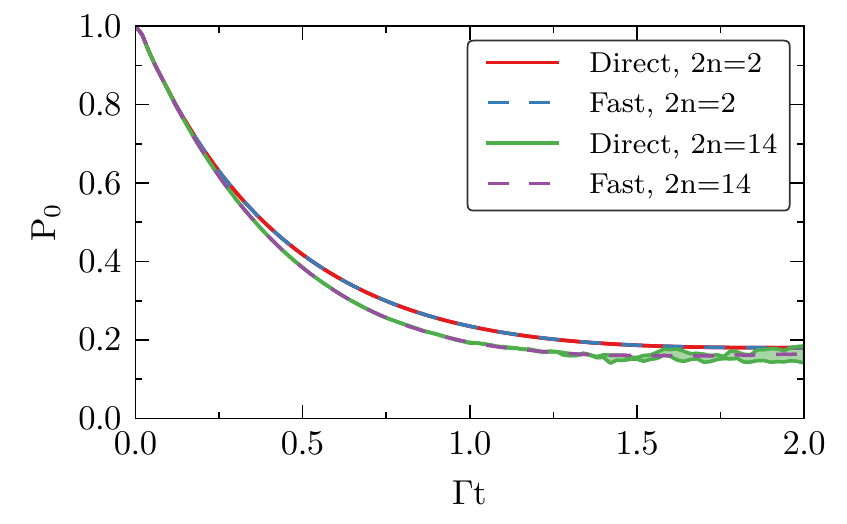}

\caption{Time dependent population of the unoccupied state in an Anderson impurity
model under a coupling quench, using the direct and optimized inclusion--exclusion
algorithms. For both algorithms, we perform calculations up to perturbation
orders of $2n=2$ and $2n=14$.\label{fig:pops_compare_algs}}
\end{figure}

In Fig.~\ref{fig:pops_compare_algs}, we plot the time dependence
of the dot's probability to be in the unoccupied state in which
it began, $P_{0}\left(t\right)$. The dynamics are evaluated using
Inchworm Monte Carlo, with the summations over $k$-connected diagrams
performed either directly (``Direct'') or by using the inclusion--exclusion
algorithm (``Fast''), using the same total amount of computer time.
The maximum order of diagrams sampled is limited to either 2 (where
the result is not converged) or 14 (where we will soon show that it
is converged). Statistical error estimates evaluated by the methods
introduced in Ref.~\onlinecite{cohen_taming_2015} are marked by
the width of the different curves. Both implementations of the method
produce the same result to within numerical accuracy, but the inclusion--exclusion
algorithm provides greatly improved accuracy at the higher order.

\begin{figure}
\includegraphics[width=8.6cm]{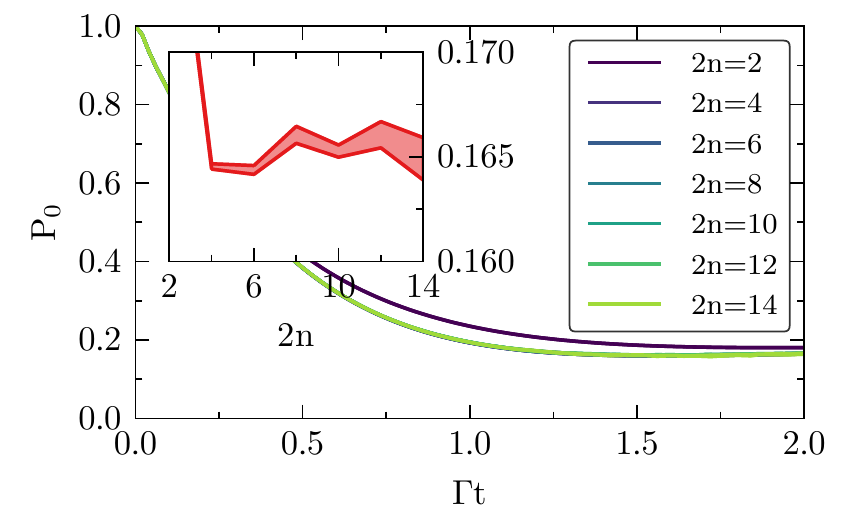}

\caption{Time dependent population of the unoccupied state in an Anderson impurity
model under a coupling quench using the optimized inclusion--exclusion
algorithm at several perturbation orders $2n$. The inset shows the
data at the final time as a function of the order, showing that 
order $2n\sim10-14$ is needed to achieve converged results at relative
errors of $\sim0.5\%$.\label{fig:pops_compare_orders}}
\end{figure}

Next, we consider convergence with the maximum diagram order. 
In Fig.~\ref{fig:pops_compare_orders}, the results from the inclusion--exclusion-based
Inchworm method are plotted at a series of maximum orders. The inset
zooms in on the result at the maximum time reached here, $\Gamma t=2$,
where it can be seen that to obtain convergence within the error bars
it is necessary to go to orders $2n\apprge12$ or 14. In this case,
convergence corresponds to relative errors of $\apprge0.5\%$ in $P_{0}$.

The computer time used to obtain each line in Figs.~\ref{fig:pops_compare_algs}
and \ref{fig:pops_compare_orders} is constant, and the errors clearly
increase with order. We now turn to studying how these errors, which
become approximately constant at long times, vary with the maximum
perturbation order. This procedure is ultimately what will determine
the usefulness of the inclusion--exclusion algorithm within the Inchworm
method: in practice, a faster summation method allows us to sample more
diagrams using the same computational resources, thus reducing the
statistical errors.

\begin{figure}
\includegraphics[width=8.6cm]{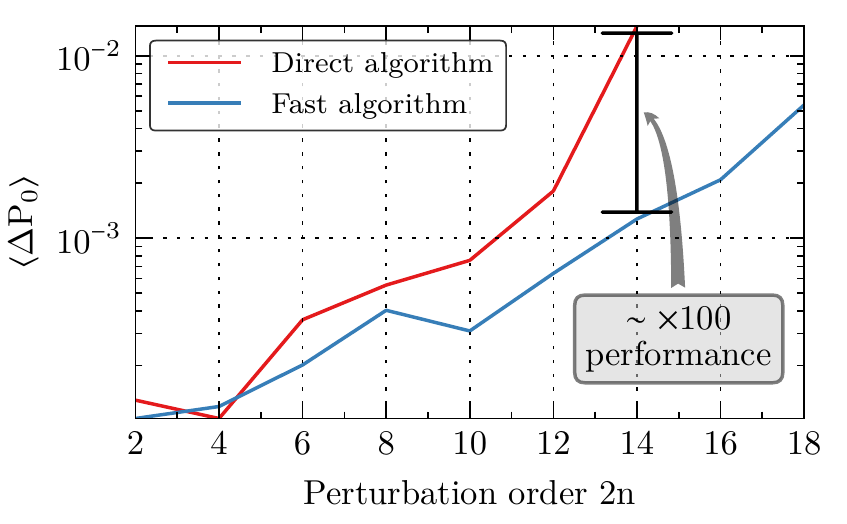}

\caption{Average errors at long times as a function of the perturbation order
$2n$, using the direct and inclusion--exclusion algorithms, for
the same parameters used in Figs.~\ref{fig:pops_compare_orders}
and \ref{fig:pops_compare_algs}. Since Monte Carlo errors scale with
the inverse square root of the computation time, an order of magnitude
reduction in the error, which is reached at $2n\apprge14$, corresponds
to a two order of magnitude enhancement in the computational efficiency.\label{fig:error_scaling}}
\end{figure}

In Fig.~\ref{fig:error_scaling}, we plot the average error at times
$1.8\le t\le2$ as a function of the maximum perturbation order $2n$,
using both algorithms. Outside a small region at $2n=4$, which is
most likely due to statistical fluctuations in our sampling, the new
algorithm is substantially faster. At the highest perturbation order
we were able to reach using the direct algorithm, $2n=14$, the inclusion--exclusion
algorithm provides errors smaller by approximately an order of magnitude
(at higher orders so few diagrams are sampled that the result becomes
unreliable without using more computer time). As errors in Monte Carlo
procedures scale with the computer time $T$ as $\frac{1}{\sqrt{T}}$,
obtaining the same reduction in error with the previous algorithm
would entail using approximately two orders of magnitude more computational
resources. At even higher orders, we expect this factor to increase
rapidly.

\section{Conclusions\label{sec:Conclusions}}

We proposed, analyzed and tested an algorithm based on the inclusion--exclusion
principle. The algorithm sums all connected (or $k$-connected) diagrams
in continuous time hybridization expansions, which are needed within
Inchworm Monte Carlo methods, in exponential time instead of the previous
factorial time. In practice, with two additional optimizations that
we proposed, the exponent we found depends to some degree on the model
in question, but if no symmetries are taken advantage of the algorithm
is $O\left(\gamma^{2n}\right)$ where $\gamma\approx1.33$ and $2n$
is the perturbation order (odd orders $2n+1$ can be ruled out for models where
the number of fermions is conserved). We also derived a rigorous upper
bound and an approximate lower bound for this exponent.

We applied the algorithm to a physical problem requiring high perturbation
orders, and showed that at reasonable parameters and accuracy it provides
a practical speedup of two orders of magnitude when compared to our
previous implementation. We note that this speedup is implementation
dependent, and we believe it can be improved even further by optimizing
parts of the code which had been of negligible computational importance
until now. However, the scaling with problem size is universal. Furthermore,
a variety of other calculations, in particular those involving Green's
functions, will greatly benefit from generalizations of the algorithm
introduced here. This will be the subject of future work.

Our algorithm is reminiscent of one which was introduced in Ref.~\onlinecite{rossi_determinant_2017}
in order to sum connected diagrams in other Monte Carlo methods based
on the interaction expansions, where the definition of connectedness
is very different. We showed that an idea along similar lines, which
we called the ``inverted'' algorithm, is correct but less efficient
than our algorithm for the hybridization expansion. We further showed
that our inclusion--exclusion algorithm can be applied to the interaction
expansion, but---at least naively---is less efficient than the inverse
algorithm in that case. As both ideas are very general in their applicability,
it will be of interest to explore their relative merits within other
expansions, methods and models in the future.

Looking forward, improving the computational efficiency of the Inchworm
method by a practical two orders of magnitude is a major step towards
making real-time Monte Carlo a viable alternative to imaginary time
techniques. We believe further improvement will stem from this work,
such as fast update schemes, and expect the inclusion--exclusion
principle to be even more beneficial in Inchworm hybridization expansions
for multiorbital impurity models. The same ideas should also be applicable
to other Inchworm expansions. The method does not generalize to bosons,
where Wick's theorem phrases the sum as a permanent rather than a
determinant---exact computation of permanents in polynomial time
is thought to be impossible.\cite{valiant_complexity_1979} On the
other hand, bosons do not suffer from a fermionic sign problem, and
Monte Carlo algorithms for summing boson diagrams work well.\cite{pollet_recent_2012}
It would therefore be of interest to consider the usefulness of the
inclusion--exclusion principle within mixed bose--fermi systems.\cite{chen_anderson-holstein_2016}
We further believe it will find applications beyond Inchworm---for
example, in the evaluation of self energies within bare hybridization
expansions, or within bold-line Monte Carlo\cite{gull_bold-line_2010,gull_numerically_2011,cohen_greens_2014,cohen_greens_2014-1}
and DiagMC techniques.\cite{prokofev_worm_1998,prokofev_bold_2007,prokofev_bold_2008,prokofev_diagrammatic_2013}

\paragraph{Acknowledgements }

We are grateful to Olga Goulko for directing our attention to Ref.~\onlinecite{rossi_determinant_2017}.
G.C. acknowledges support by the Israel Science Foundation (Grant
No. 1604/16). E.G. was supported by DOE ER 46932.
This research was supported by Grant No.
2016087 from the United States-Israel Binational Science Foundation
(BSF).

 \bibliographystyle{apsrev4-1}
\bibliography{Library}

\appendix

\section{Derivation\label{sec:Derivation}}

In this appendix, we will introduce a precise phrasing of the celebrated
inclusion--exclusion principle, and show how it can be used to derive
Eq.~\ref{fig:Basic-algorithm}. This principle is most often stated
in terms of counting the size of a union. For example,consider two
sets $A$ and $B$. The size of their union can be written 
\begin{equation}
\left|A\cup B\right|=\left|A\right|+\left|B\right|-\left|A\cap B\right|.
\end{equation}
However, if one is given three sets $A$, $B$ and $C$, the union
is: 
\begin{equation}
\begin{aligned}\left|A\cup B\cup C\right| & =\left|A\right|+\left|B\right|+\left|C\right|-\left|A\cap B\right|\\
 & -\left|A\cap C\right|-\left|B\cap C\right|+\left|A\cap B\cap C\right|,
\end{aligned}
\end{equation}
and for $N$ sets $S_{i}$, one can write a general expression in
the form 
\begin{equation}
\begin{aligned}\left|\cup_{i=1}^{N}S_{i}\right| & =\sum_{i_{1}}\left|S_{i_{1}}\right|-\sum_{i_{1}<i_{2}}\left|S_{i}\cap S_{j}\right|\\
 & +\cdots\sum_{i_{1}<\cdots<i_{N}}\left(-1\right)^{N-1}\left|S_{i_{1}}\cap\cdots\cap S_{i_{n}}\right|.
\end{aligned}
\end{equation}
The inclusion--exclusion principle has a long history and many uses in combinatorics. Interestingly, it has also found applications in the context of nonlocal extensions to dynamical mean theory,\cite{schiller_systematic_1995,hettler_nonlocal_1998,lan_generalized_2017,zgid_finite_2017,vucicevic_practical_2018} though these works did not explicitly call it by this name. 

Here, we use a trivial generalization from the size of the sets to
a generic scalar property. The inclusion--exclusion principle as
we use it deals with a set $S$, a collection of subsets thereof $\left\{ A_{i}\right\} $,
and a function $f\colon s\in S\rightarrow\mathbb{C}$. It states that
the sum of $f$ over elements of $S$ that are not the elements of
any $A_{i}$ can be computed by first taking the sum over the values
of $f$ for all elements of $S$, then subtracting the sums of values
of $f$ for all subsets $A_{i}$, then adding the values of $f$ for
those elements which we have subtracted twice (the elements of all
sets $A_{i}\cap A_{j}$), and so on. This leads to the following equality:

\begin{equation}
\begin{aligned}f & \equiv\sum_{x\in S\backslash\left(\cup_{i}A_{i}\right)}f\left(x\right)=\sum_{x\in S}f\left(x\right)-\sum_{i}\sum_{x\in A_{i}}f\left(x\right)\\
 & +\sum_{i<j}\sum_{x\in A_{i}\cap A_{j}}f\left(x\right)-\dots\\
 & =\sum_{j=0}^{j_{\mathrm{max}}}\left(-1\right)^{j}\sum_{i_{0}<\cdots<i_{j-1}}\sum_{x\in A_{i_{0}}\cap\dots\cap A_{i_{j-1}}}f\left(x\right).
\end{aligned}
\end{equation}

To obtain our algorithm, we set $D$ to be the set of all diagrams
over vertices $S$ and $f$ to be the function that associates values
with diagrams. We define $A_{S}$ to be the set of connected diagrams
over $S$, and obtain $\left\{ A_{s}\right\} $ for every $j$ by
collecting all sets of $j$ disjoint subsegments of $S$ comprising
only proper vertices. Since (a) every diagram which isn't $k$-connected
has a connected fully proper segment; and (b) sets of connected segments
are necessarily disjoint, this leaves only proper diagrams. Applying
the inclusion--exclusion principle we immediately get Eq.~(\ref{eq:algorithm_naive}),
using the fact that the sum over values of diagrams for which the
segments $S_{0},\dots,S_{j-1}$ are connected is $c\left(S_{0}\right)\cdots c\left(S_{j-1}\right)v\left(V\backslash\cup_{i=0}^{j-1}C\left(s_{i}\right)\right)$.

\section{Computational efficiency\label{sec:Computational-efficiency}}

In this appendix, we show how the theoretical bounds $\alpha$ and
$\beta$ can be derived for the different approximations discussed
above. A fully analytical combinatorial computation is possible, but
lengthy. Since we are interested only in the asymptotic scaling, we
will limit ourselves to scaling calculations based on the pole structure
of the generating functions of the relevant sequences.

\subsection{Upper bound}

To simplify the derivation, we will ignore the distinction between
creation and annihilation operators and analyze the complexity of
the generic algorithm in which every vertex can be paired to other
vertex. Of course, it is possible to implement the physical expansion
with this algorithm by setting elements of of $M_{ij}$ between operators
of the same type to zero. However, the number of diagrams that needs
to be summed is exponentially larger and it is clear that this will
provide an overestimate of the fermionic algorithm; the result will
therefore be useful as an upper bound.

Throughout this subsection, we will assume that $V$ is the set of
proper vertices, of size $m$, and that there exist $k$ additional
improper vertices. In the physical case, one would have $m=2n-k$.

\subsubsection{Unoptimized algorithm}

Consider first how Eq.~(\ref{eq:algorithm_naive}) is used in practice:
we must compute $c\left(S\right)$ for all segments $S\subset V$
in increasing order of size, as each segment will depend on results
involving smaller segments. Finally, $c\left(V\right)$ will be evaluated.
Each step takes a number of evaluations of $v$ equal to the number
of ways to choose sets of disjoint segments. Let $a_{m}$ denote this
number for a set of size $m$ The complexity of each step is then
$O\left(\left(m+p\right)^{3}a_{m}\right)$, since the most expensive
part for each set of segments is the evaluation of $v$. Since the
number of steps of each size smaller than the final $m$ is polynomial,
while $a_{m}$ turns out to be exponential in $m$, the final step
dominates the complexity.

We now continue to the combinatorial calculation of $a_{m}$. Each
set of segments either contains a segment including the last point,
or does not. If it does, and this segment is of length $\ell$ (which
is even, as in a subset of odd length not all vertices can be paired),
then we are left with $a_{m-\ell}$ options to choose the rest of
the subsets. If it has no segment including the last point, there
are $a_{m-1}$ options. Therefore, we get 
\begin{equation}
a_{m}=a_{m-1}+a_{m-2}+a_{m-4}+\dots+a_0\label{eq:rec_relation_noops}
\end{equation}
for any $m>0$, and for convenience we set $a_{0}=1$.

To find the asymptotic growth rate of the sequence, it is useful to
consider its generating function $f\left(x\right)\equiv\sum_{m=0}^{\infty}a_{m}x^{m}$.
Using Eq.~(\ref{eq:rec_relation_noops}), 
\begin{equation}
\sum_{m=1}^{\infty}\left(a_{m}-a_{m-1}-a_{m-2}-a_{m-4}-\dots\right)x^{m}=0,
\end{equation}
or 
\begin{equation}
f\left(x\right)-a_{0}-\left(x+x^{2}+x^{4}+\ldots\right)f\left(x\right)=0.
\end{equation}
Using our value for $a_{0}$ and summing the series, we obtain 
\begin{equation}
\left(1-x-\frac{x^{2}}{1-x^{2}}\right)f\left(x\right)=1,
\end{equation}
such that 
\begin{equation}
f\left(x\right)=\frac{1-x^{2}}{1-x-2x^{2}+x^{3}}.
\end{equation}
If asymptotically $a_{m}\sim\alpha^{m}$, $f\left(x\right)$ will
have a pole with absolute value $\frac{1}{\alpha}$ and no poles with
smaller absolute value. The smallest pole by absolute value is at
$\left|x_{\mathrm{min}}\right|\approx0.55495$, such that $\alpha=\frac{1}{\left|x_{\mathrm{min}}\right|}\approx1.8019$.

\subsubsection{Effect of optimizations}

We can obtain an analogous formula for the number $a_{m}$ of ways
to choose non-adjacent disjoint subsegments of $m$ vertices by considering
three options at every stage: (a) there is no segment containing the
last point, giving $a_{m-1}$ possible choices; (b) there is a segment
of length $\ell$ containing the last point, before which there is
a vertex which is not an element of any segment, giving $a_{m-2\ell-1}$
choices; and (c) there is a single subsegment that contains all vertices,
giving 1 option if $m$ is even. Therefore, 
\begin{equation}
a_{m}=\sum_{\ell=0}^{m/2}a_{m-2\ell-1}+\left(1\text{ if \ensuremath{m} is even}\right).
\end{equation}
As before, we can show that the generating function $f\left(x\right)$
for this sequence satisfies

\begin{equation}
f\left(x\right)=\frac{1}{1-x-x^{2}},
\end{equation}
for which the growth rate is the golden ratio $\alpha\approx1.6180$.

After the second optimization, we need not count segments of size
2. By analogous considerations this gives a sequence generated by
\begin{equation}
f\left(x\right)=\frac{1-x^{2}+x^{4}}{1-x-x^{2}+x^{3}-x^{5}},
\end{equation}
which has the growth rate $\alpha\approx1.4432$.

\subsubsection{Inverted algorithm}

To evaluate the performance of the algorithm implied by Eq.~\ref{eq:inverted_algorithm},
we need to calculate the number of ways to choose sets of subsegments
containing a total of $\ell$ vertices, from a set with $n$ vertices.
We will denote this number by $s_{m}^{\ell}$. Given that, the runtime
of the algorithm is given by $r_{m}=\sum_{\ell\le m}s_{m}^{\ell}a_{\ell}$,
where $a_{\ell}$ is the same runtime we evaluated for the inclusion--exclusion
algorithm with the first optimization. Since we only care about the
asymptotic growth rate, and we found that $a_{\ell}=O\left(\beta^{\ell}\right)$,
then

\begin{equation}
r_{m}=O\left(\sum_{\ell\le m}s_{m}^{\ell}\beta^{\ell}\right).
\end{equation}
Let $S\left(x,\beta\right)\equiv\sum_{m,\ell}s_{m}^{\ell}x^{m}\beta^{\ell}$
be the generating function of $s_{m}^{\ell}$. The growth rate of
$r_{m}$ is therefore given by that of the coefficient of $x^{m}$
in $S\left(x,\beta\right)$.

Therefore, we are only left with the task of evaluating $S\left(x,t\right)$.
By similar arguments to those used before, $s_{n}^{\ell}$ satisfies

\begin{equation}
s_{m}^{\ell}=\sum_{j}\left(s_{m-2j-1}^{\ell-2j}+\left(1\text{ if \ensuremath{m=\ell=2j}}\right)\right),
\end{equation}
so

\begin{equation}
S\left(x,\beta\right)=\frac{1}{1-x-x^{2}\beta^{2}}.
\end{equation}
Substituting $\beta\approx1.8019$ from the first optimization case
and repeating the procedure from before, we get that $\alpha\approx2.1935$.
Similarly, with the second optimization, $\alpha\approx1.8718$.

\subsection{Average over operator orders}

We will now consider a case closer to the one which is of physical
interest, by taking into account the fact that vertices consist of
either creation or annihilation operators and that pairing can only
occur between operators of different type. This is a more complex
calculation, and we will only show how it is performed for the unoptimized
case. The effect of the optimizations and of the growth rate of the
inverted algorithm can be similarly calculated.

To address this case, we will first count the number of ways $a_{n,n}$
to partition $2n$ vertices into two subsets of $n$ vertices corresponding
to creation and annihilation operators and then choose subsegments
of the full set containing the same number of operators of each type.
To perform the averaging over the possible operator orders, we will
then divide this result by the number of ways to partition the operators
into the two types, $\binom{2n}{n}$. Since $a_{n,n}$ will turn out
to be exponential in $n$ and $\binom{2n}{n}\simeq\frac{4^{n}}{\sqrt{n}}$,
the growth rate for the average will be the growth rate of $a_{n,n}$
divided by $4$.

It turns out that it is easier to solve a slightly more general combinatorial
problem: the number of ways $a_{m,n}$ to partition $m+n$ vertices
into two subsets, one containing $m$ vertices and the other containing
$n$ vertices, and then choose subsegments accordingly. This obeys
the following recurrence relation:

\begin{equation}
a_{m,n}=a_{m-1,n}+a_{m,n-1}+\sum_{j=1}^{m}\binom{2j}{j}a_{m-j,n-j},
\end{equation}
for $m,\thinspace n>0$ and $a_{0,0}=1$. Therefore, the generating
function of $a_{m,n}$, $g\left(x,y\right)\equiv\sum_{m,n}a_{m,n}x^{m}y^{n}$,
satisfies the equation

\begin{equation}
\begin{aligned}g\left(x,y\right) & =\left(x+y\right)g\left(x,y\right)\\
 & +\left(\frac{1}{\sqrt{1-4xy}}-1\right)g\left(x,y\right)+1,
\end{aligned}
\end{equation}
where we have used the fact that

\begin{equation}
\sum_{k=0}^{\infty}\binom{2k}{k}x^{k}=\frac{1}{\sqrt{1-4x}}.
\end{equation}
Solving this, we obtain 
\begin{equation}
\begin{aligned}g\left(x,y\right) & =\frac{1}{2-x-y-\left(1-4xy\right)^{-\frac{1}{2}}}\\
 & =\frac{1}{2-\left(1-4xy\right)^{-\frac{1}{2}}}\cdot\frac{1}{1-\frac{x+y}{2-\left(1-4xy\right)^{-\frac{1}{2}}}}.
\end{aligned}
\label{eq:avg_alg_noopt_genfunc}
\end{equation}

We are actually interested in the sequence $a_{n,n}$, and its generating
function $f\left(x\right)=\sum_{n}a_{n,n}x^{n}$. However, $f\left(xy\right)$
contains the terms of $g\left(x,y\right)$ which have the same power
of $x$ and $y$. Since the only term in $g\left(x,y\right)$ that
can contribute differently in Eq.~(\ref{eq:avg_alg_noopt_genfunc})
is $x+y$, we can expand the second fraction in a series: 
\begin{equation}
\begin{aligned}\frac{1}{1-\frac{x+y}{2-\left(1-4xy\right)^{-\frac{1}{2}}}} & =\sum_{k=0}^{\infty}\left(\frac{x+y}{2-\left(1-4xy\right)^{-\frac{1}{2}}}\right)^{k}\\
 & =\sum_{k=0}^{\infty}\left(2-\left(1-4xy\right)^{-\frac{1}{2}}\right)^{-k}\\
 & \times\sum_{i=0}^{k}\left(\begin{array}{c}
i\\
j
\end{array}\right)x^{i}y^{k-i}.
\end{aligned}
\end{equation}
As we only want the terms with equal powers of $x$ and $y$, we need
only take the terms with $i=k-i$, \emph{i.e.} $k=2i$. With this,

\begin{equation}
\begin{aligned}f\left(xy\right) & =\frac{1}{2-\left(1-4xy\right)^{-\frac{1}{2}}}\\
 & \times\sum_{i=0}^{\infty}\left(2-\left(1-4xy\right)^{-\frac{1}{2}}\right)^{-2i}\left(\begin{array}{c}
2i\\
i
\end{array}\right)\left(xy\right)^{i},
\end{aligned}
\end{equation}
from which we can obtain 
\begin{equation}
f\left(x\right)=\left(\left(2-\left(1-4x\right)^{-\frac{1}{2}}\right)^{2}-4x\right)^{-\frac{1}{2}}.
\end{equation}
Finally, substituting $x=\frac{1}{4\beta^{2}}$, we get that the growth
rate is the largest solution of $\frac{1}{\beta}=2-\left(1-\frac{1}{\beta^{2}}\right)^{-\frac{1}{2}}$,
such that $\beta\approx1.5072$. 
\end{document}